\documentclass[3p,times,twocolumn]{elsarticle}

\usepackage{ecrc}
\usepackage{amsmath}

\volume{00}

\firstpage{1}

\journalname{Nuclear and Particle Physics Proceedings}

\runauth{}

\jid{nppp}

\jnltitlelogo{Nuclear and Particle Physics Proceedings}

\usepackage{amssymb}

\usepackage[figuresright]{rotating}

\begin{document}
\begin{frontmatter}

\dochead{}

\title{Measurement of $D^{*}$-meson triggered correlations in p+p collisions at RHIC}

\author{Long Ma$^{1,2}$\fnref{ax} for the STAR Collaboration}

\address{$^{1}$Shanghai Institute of Applied Physics, Chinese Academy of Sciences, Shanghai 201800, China \\
$^{2}$University of Chinese Academy of Sciences, Beijing 100049, China}
\fntext[ax]{Email address: malong@sinap.ac.cn}

\begin{abstract}

We report the preliminary results of the azimuthal correlations between $D^{*\pm}$ mesons and charged hadrons ($D^{*}$-h) measured by the STAR experiment in proton+proton collisions at $\sqrt{s}$ = 500 GeV. Results at mid-rapidity in the transverse-momentum range 8 $\le$ $p^{D^{*}}_{T}$ $\le$ 20 GeV/$c$ are compared with light hadron triggered correlations (h-h) and PYTHIA predictions. We also present an exploratory study of azimuthal correlations between $D^{*+}$ and $D^{*-}$ mesons in p+p collisions. The prospects of measuring heavy-flavor triggered correlations in heavy-ion collisions at RHIC energies are also discussed.

\end{abstract}


\end{frontmatter}


\section{Introduction}
\label{intro}

At RHIC energies, heavy quarks are mostly produced in pairs through initial hard scatterings, early before the formation of the strongly interacting medium - the Quark-Gluon Plasma (QGP). Primordial heavy quarks experience the entire evolution of the medium and carry important information about the QGP. These features make heavy quarks a crucial probe to study the QGP properties~\cite{ref1,ref2,ref3,ref4}. 

Measurements of angular correlations triggered by heavy quarks in heavy-ion collisions can provide insights into the energy loss mechanism of heavy quarks in the QGP~\cite{ref7,ref8}. Recent studies also suggest that azimuthal correlations between heavy quark pairs uniquely probe the heavy quark-medium interaction dynamics, and therefore has the potential for distinguishing different energy loss mechanisms for heavy quarks~\cite{ref9,ref10,ref11}. 

In p+p collisions, measurements of heavy-flavor triggered correlations allow to test perturbative QCD (pQCD) calculations and provide references for studies in heavy-ion collisions. Direct comparison between heavy-flavor and light-flavor triggered correlations can shed lights on the differences in fragmentation and hadronization between heavy and light quarks~\cite{ref12,ref13}.

\section{$D^{*}$-meson triggered angular correlations}
\label{dh}

The analysis of $D^{*}$-meson triggered angular correlations is carried out using a data set of about 160 million p+p collisions at $\sqrt{s}$ = 500 GeV. These events are required to contain high-energy depositions exceeding 4.2 GeV in the Barrel Electromagnetic Calorimeter (BEMC). $D^{*\pm}$ mesons are reconstructed via their hadronic decay channels, i.e. $D^{*\pm}\rightarrow D^{0}\pi ^{+}(\bar{D}^{0}\pi ^{-})$ with a branching ratio of 67.7$\%$ and $D^{0}(\bar{D}^{0})\rightarrow K^{-}\pi ^{+}(K^{+}\pi ^{-})$ with a branching ratio of 3.89$\%$. Specific energy loss (dE/dx) measured by the Time Projection Chamber (TPC) allows for identification of charged particles at mid-rapidity~\cite{ref14}. The Time Of Flight (TOF) detector is also used to identify daughter kaons in case the time-of-flight information is available. General track quality cuts are applied on all the charged tracks at mid-rapidity ($|\eta|$ $<$ 1) and the transverse momenta ($p_{T}$) of the associated hadrons are required to be above 0.5 GeV/$c$.

\begin{figure}[!htb]
\centering
\includegraphics[width=0.4\textwidth,height=0.3\textwidth]{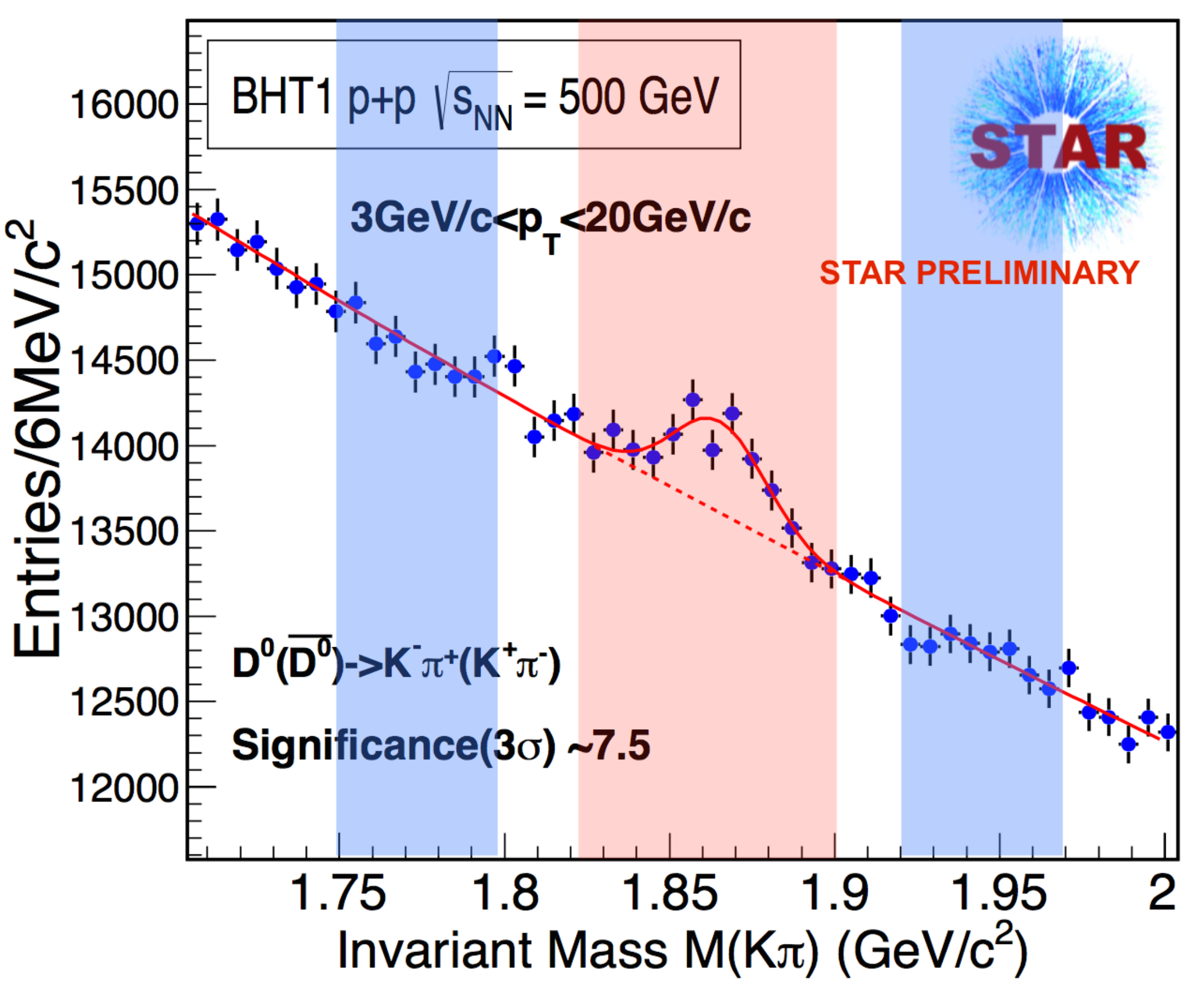}\vspace{0.2cm}
\includegraphics[width=0.4\textwidth,height=0.3\textwidth]{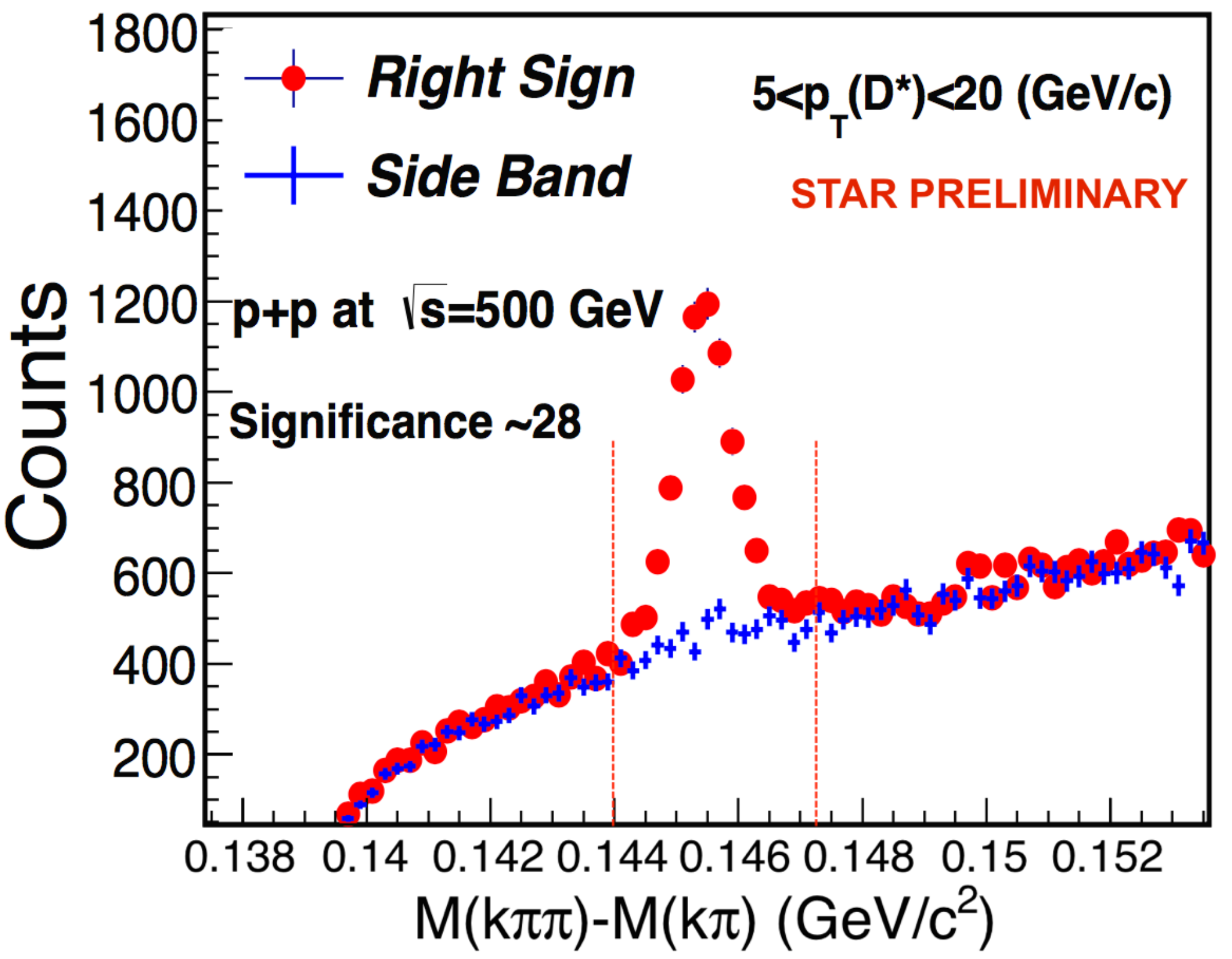}
\caption{(Color online) $D$ meson signal at mid-rapidity in p+p collisions at $\sqrt{s}$ = 500 GeV. (Top panel) Invariant mass distribution of opposite-sign $K\pi$ pairs. The solid red curve shows the combined function fit to the mass distribution. The $D^0$ signal and side-band background regions are shaded with red and blue, respectively. (Bottom panel) Invariant mass difference $\Delta M=M(K\pi\pi)-M(K\pi)$ distribution for $D^{*\pm}$ candidates (in red). The blue data points represent combinatorial background reconstructed using the $K\pi$ pairs in the side-band regions shown in the upper panel. Candidates within $3\sigma$ window of the signal peak (red dashed lines) are selected for the correlation study.}
\centering
\label{f1}
\end{figure}

The top panel of Fig.~\ref{f1} shows the invariant mass distribution of opposite-sign $K\pi$ pairs. $D^{0}$ candidates within $\pm$3$\sigma$ of the signal peak (red area), where $\sigma$ is the width of the peak, are selected for reconstructing $D^{*\pm}$ mesons. $K\pi$ pairs with invariant mass outside the $D^{0}$ invariant mass region (side-band, blue areas) are selected to reconstruct $D^{*\pm}$ background. The side-band background is scaled to match the background yield under the $D^{0}$ signal peak. $D^{*\pm}$ mesons are reconstructed by combining one low momentum pion with one $D^{0}$ candidate. The side-band background well reproduces the combinatorial background for $D^{*\pm}$ as shown in the bottom panel of Fig.~\ref{f1}. 

Reconstructed $D^{*\pm}$ mesons are selected within $\pm$3$\sigma$ from the center of the signal peak, and correlated with charged hadrons in the same event with triggered $D^{*\pm}$ decay daughters excluded. The resulting angular correlations contain contributions from both signal and background. Since events used in this analysis are triggered by the BEMC, $D^{0}$-decayed daughter pions are required to fire the trigger and the trigger inefficiency is corrected accordingly. Associated charged hadrons are required to match one of the fast detectors, i.e. the TOF or BEMC, to reduce the pile-up effect. Efficiency corrections are applied for both $D^{*\pm}$ and associated charged hadrons. Additional effects from detector inhomogeneities and limited detector acceptance are also accounted for.

\begin{figure}[!htb]
\centering
\includegraphics[width=0.45\textwidth]{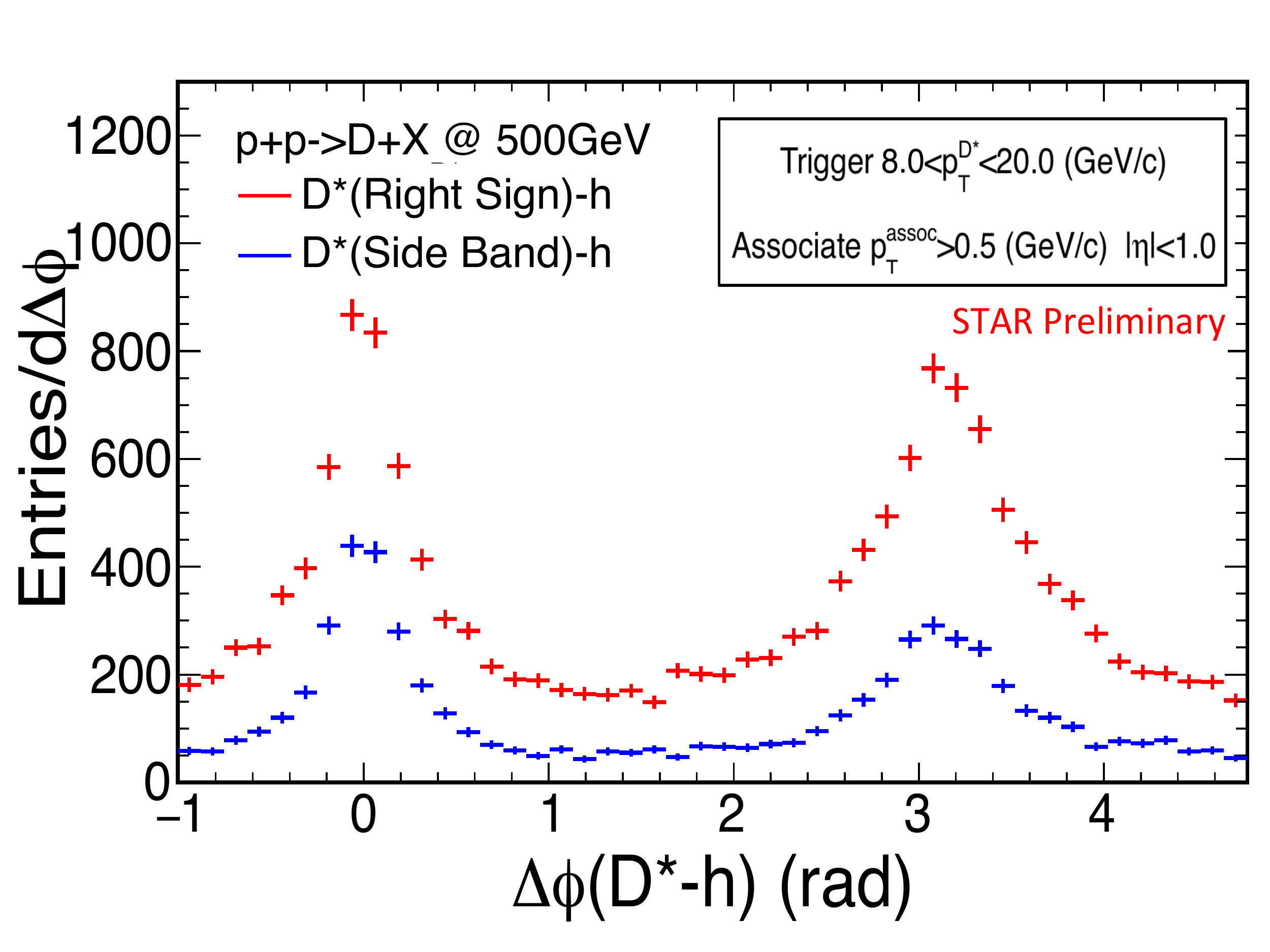}
\caption{(Color online) Comparison of azimuthal correlations triggered by $D^{*\pm}$ candidates (including both signal and background) and side-band background with charged hadrons in p+p collisions at $\sqrt{s}$ = 500 GeV. The red points are for $D^{*\pm}$ candidates, while the blue points are for side-band background.}
\centering
\label{f2}
\end{figure}

Fig.~\ref{f2} shows the angular correlations triggered by $D^{*\pm}$ candidates (``Right Sign", red points) and background (``Side Band", blue points) for $8 < p_{T}^{Trig} < 20$ GeV/$c$ and $p_{T}^{assoc} > 0.5 $ GeV/$c$.

\section{Result and discussion}
\label{result}

The correlation signal is obtained by subtracting the side-band background triggered azimuthal correlations from those triggered by $D^{*\pm}$ candidates, where the side-band background is scaled to match the background yield under the $D^{*\pm}$ peak. More specifically

\begin{equation}
C_{signal}(\Delta \varphi,|\eta|<1) = C(\Delta \varphi,|\eta|<1)_{RS} - \frac{BG_{RS}}{BG_{S\!B}}C(\Delta \varphi,|\eta|<1)_{S\!B}
\label{q1}
\end{equation}

where $D^{*\pm}$ candidates are denoted as ``RS" and ``SB" stands for side-band background. The resulting azimuthal correlations are then normalized by the total number of $D^{*\pm}$ mesons, i.e. $1/N_{trig}dN_{asso}/d\Delta\phi$, and fitted with a double-Gaussian function on top of a constant baseline 

\begin{equation}
\begin{split}
& f(\Delta \phi ) = \frac{Y_{NS}}{\sqrt{2\pi }\sigma _{NS}}exp(-\frac{(\Delta \phi_{NS})^{2}}{2\sigma^{2}_{NS} }) \\ 
& +\frac{Y_{AS}}{\sqrt{2\pi }\sigma _{AS}}exp(-\frac{(\Delta \phi_{AS})^{2}}{2\sigma^{2}_{AS} })+baseline \\
\end{split}
\label{q2}
\end{equation}

The associated yields ($Y_{NS}$ for the near-side and $Y_{AS}$ for the away-side) and the baseline originating from the uncorrelated background can be extracted from the fitting.

\begin{figure}[!htb]
\centering
\includegraphics[width=0.45\textwidth]{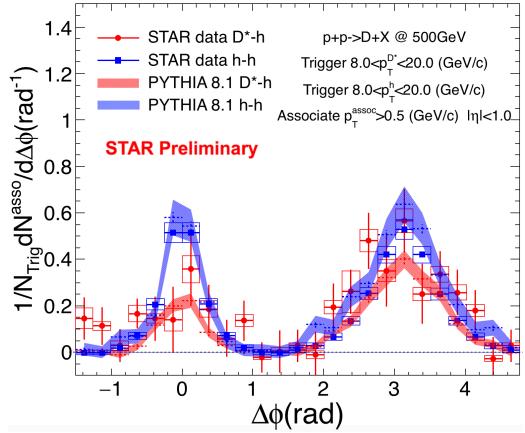}
\caption{(Color online) Azimuthal correlations between $D^{*\pm}$ mesons and charged hadrons in p+p collisions after baseline subtraction. $D^{*}$-hadron correlations (red solid circles) for 8 $<$ $p^{Trig}_{T}$ $<$ 20 GeV/$c$ are compared to similar correlations triggered by light hadrons in the same kinematic region (blue solid squares). PYTHIA predictions (shaded areas) are also shown for comparison.}
\centering
\label{f3}
\end{figure}

 Fig.~\ref{f3} shows the corrected azimuthal correlations between $D^{*\pm}$ mesons and charged hadrons within $|\eta|$ $<$ 1 for 8 $<$ $p_{T}^{Trig}$ $<$ 20 GeV/$c$ and $p_{T}^{assoc}$ $>$ 0.5 GeV/$c$. PYTHIA 8.1 is tuned to reproduce the inclusive charm quark production cross-section in p+p collisions at $\sqrt{s}$ = 500 GeV measured by the STAR experiment, and then used to estimate the feed-down contribution from B-hadron decays which turns out to be around 5$\%$~\cite{ref15,ref16,ref17,ref18}.  

Comparisons are made between $D^{*}$-hadron and hadron-hadron correlations with baseline subtracted. The baseline variation is included in the systematic uncertainties. As shown in Fig.~\ref{f3}, the away side ($\Delta\phi \sim \pi$) of the $D^{*}$-hadron correlations agrees with light hadron triggerd correlations within uncertainties. Better statistical precision is needed for the near-side comparison. Both $D^{*}$-hadron and hadron-hadron correlations are compared to PYTHIA predictions, and they are found to be compatible within uncertainties.

The azimuthal correlations between $D^{*+}$ and $D^{*-}$, which are expected to access the correlations between parent charm and anti-charm quarks, are studied using the same data set. The correlation between $D^{*+}$ and $D^{*-}$ is derived as 
  
\begin{equation}
\begin{aligned}
C^{D\overline{D}} = RS(D^{*+}) \ast RS(D^{*-}) - S\!B(D^{*+}) \ast RS(D^{*-})\\
- S\!B(D^{*-}) \ast RS(D^{*+}) + S\!B(D^{*+}) \ast S\!B(D^{*-})
\end{aligned}
\label{q3}
\end{equation}

where the asterisks (*) indicate the correlations between the two terms. This background subtraction method has been verified with PYTHIA simulations, and the corrected correlations can well reproduce the true correlations.

\begin{figure}[!htb]
\includegraphics[width=0.4\textwidth]{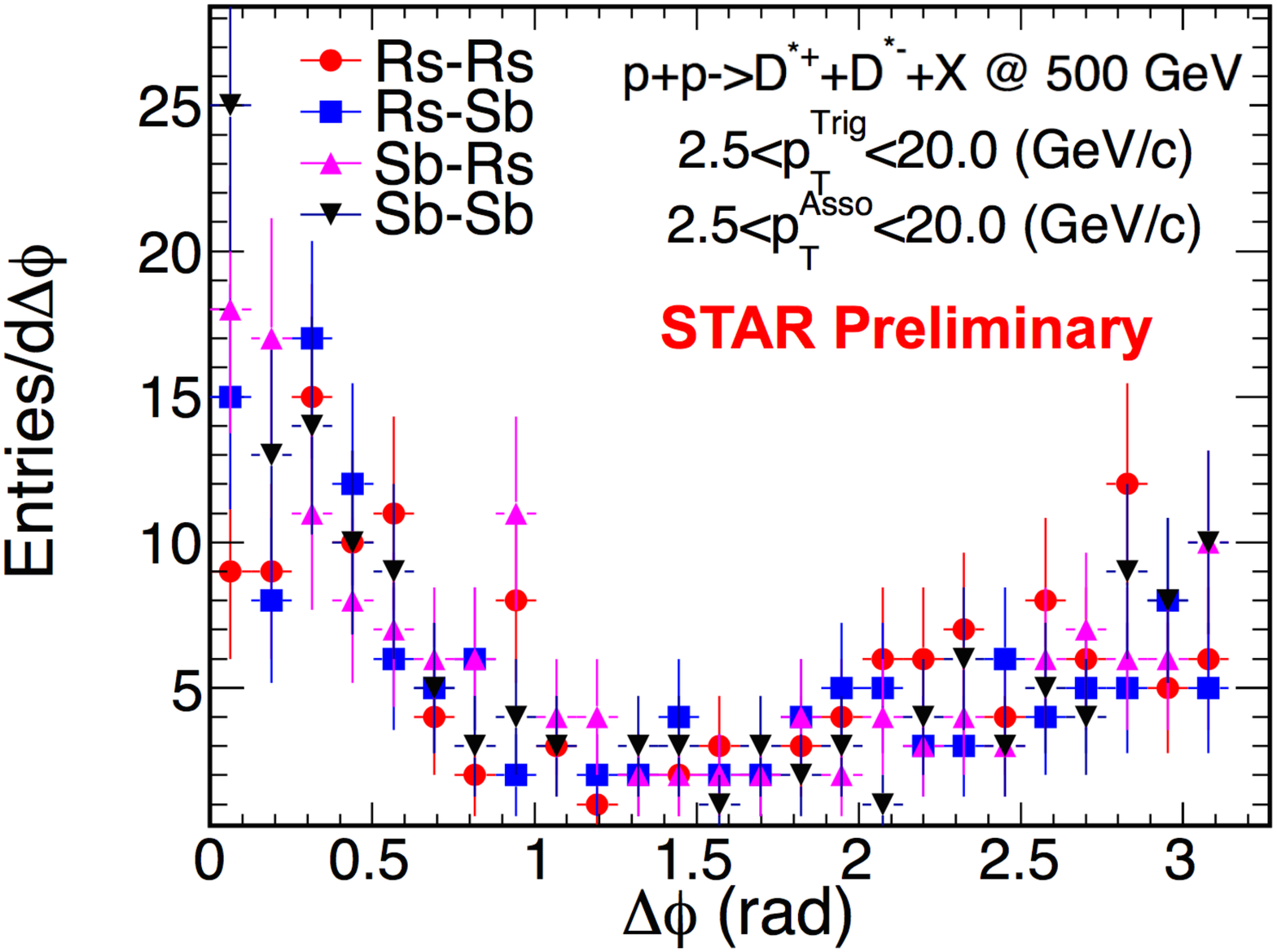}
\includegraphics[width=0.4\textwidth]{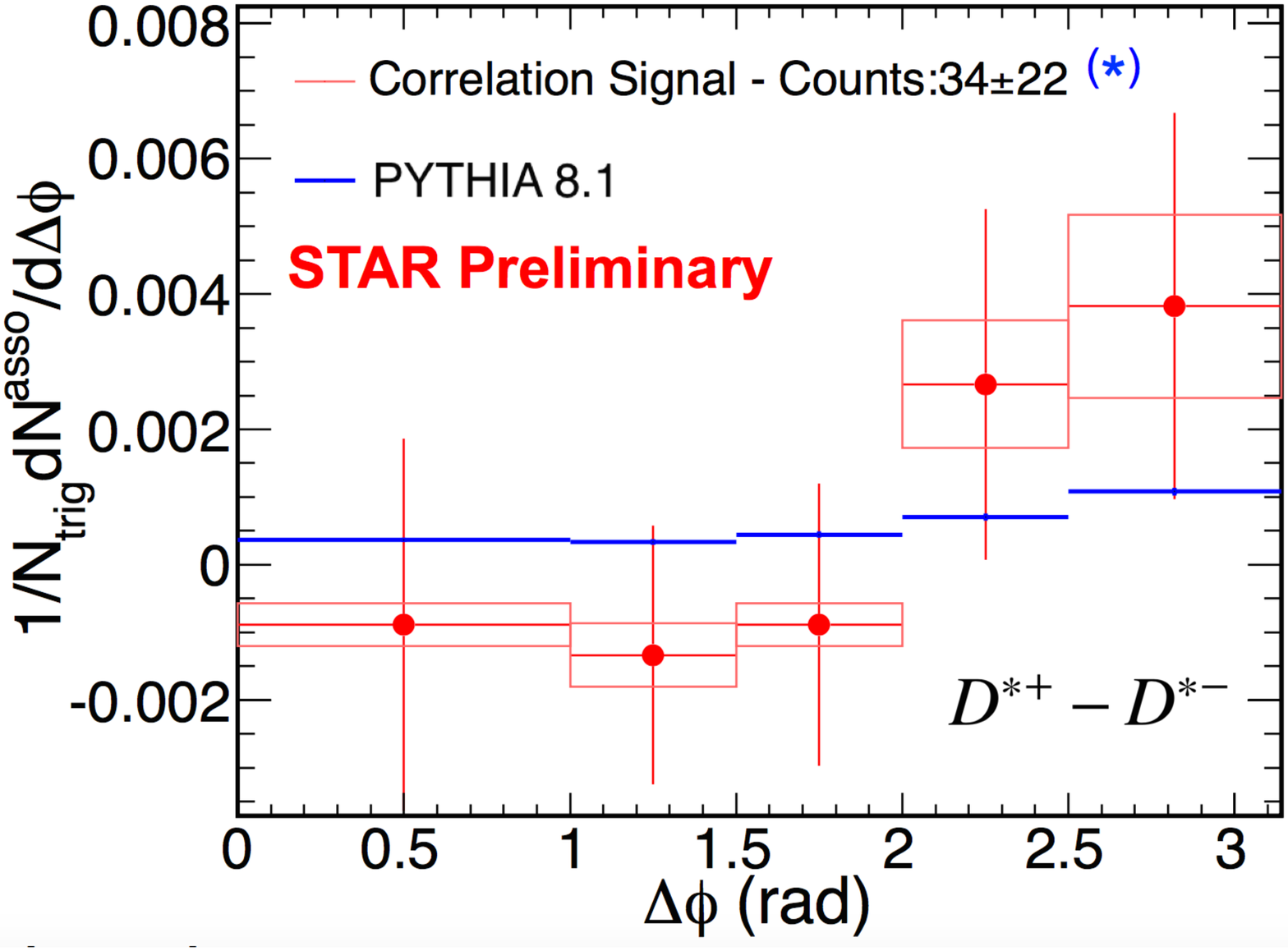}
\caption{(Color online) Azimuthal correlations between $D^{*+}$ and $D^{*-}$ in p+p collisions at $\sqrt{s}$ = 500 GeV. (Upper panel) Cross correlations of different combinations of $D^{*\pm}$ candidates (RS) and side-band background (SB). Both $D^{*+}$ and $D^{*-}$ are required to have $p_{T}$ above 2.5 GeV/$c$. (Lower panel) Preliminary results of $D^{*+}-D^{*-}$ correlation after background subtraction based on Eq.~\ref{q3}.}
\label{f10}
\end{figure}
 
In Fig.~\ref{f10}, the upper panel shows correlations as a function of $\Delta \phi$ between different combinations of signal candidates and background. The $p_T$ $>$ 2.5 GeV/$c$ cut is applied for both $D^{*+}$ and $D^{*-}$ mesons. The $D^{*+}-D^{*-}$ correlations after background subtraction based on Eq.~\ref{q3} and normalization by the number of triggers are shown in the lower panel of Fig.~\ref{f10}. A total number of 34$\pm$22 correlated $D^{*+}-D^{*-}$ pairs is observed. A PYTHIA prediction using the same tune as in the $D^{*}$-h correlation study is also shown in Fig.~\ref{f10}, and it is compatible with data within uncertainties.

\section{Conclusions}
\label{summary}

We report the study of the heavy-quark triggered correlations via the measurement of azimuthal correlations between $D^{*\pm}$ mesons and charged hadrons at mid-rapidity in p+p collisions at $\sqrt{s}$ = 500 GeV. Results are compared with light-hadron triggered correlations and PYTHIA predictions. The away side of the $D^{*}$-hadron correlations is compatible with that of light-hadron triggered correlations for $8 < p_{T}^{Trig} < 20$ GeV/$c$ and $p_{T}^{assoc} > 0.5$ GeV/$c$ within uncertainties. The first measurement of $D^{*+}-D^{*-}$ correlations in 500 GeV p+p collisions is also presented, which is found to be consistent with PYTHIA predictions within uncertainties.

Experimental studies of $D$-hadron or $D-\overline{D}$ azimuthal correlations are challenging in heavy-ion collisions because charm hadron reconstructions through their hadronic decay channels suffer from overwhelming background.

In 2014, STAR had its inner tracking system upgraded, i.e. the Heavy Flavor Tracker (HFT), installed and commissioned. Significantly improved track pointing resolution enables $D$ meson reconstruction with high signal-to-background ratio. High-statistics data sets taken with the HFT in the RHIC 2014 and 2016 runs provide a unique opportunity to study heavy-flavor triggered correlations in Au+Au collisions in near future.

\section*{Acknowledgement}
\label{ack}

This work was supported in part by the National Natural Science Foundation of China (Grant No.11421505 and No.11220101005) and the Major State Basic Research Development Program in China (Grant No.2014CB845400).

\end{document}